\documentclass[aip,pop,reprint,amsmath,amssymb,nofootinbib]{revtex4-1}

\usepackage{geometry}
\usepackage{color}
\usepackage{calc}
\usepackage{amsmath}
\usepackage{amssymb}
\usepackage{cancel}
\usepackage{graphicx} 
\usepackage{esint}
\PassOptionsToPackage{normalem}{ulem}
\usepackage{ulem}
\usepackage{sidecap}
\makeatletter

\usepackage{bm}
\usepackage{soul}
\usepackage{url}
\usepackage{varioref}
\usepackage{graphics}
\usepackage{subfigure}
\usepackage{adjustbox}
\usepackage{appendix}
\usepackage{multirow}
\usepackage{verbatim}
\usepackage{colortbl}
\usepackage{times}
\usepackage{multimedia}
\usepackage{enumerate}
\usepackage{dcolumn}
\usepackage[usenames,dvipsnames,svgnames,table]{xcolor}
\usepackage{wrapfig}
\usepackage{nth}

\newcommand{\mf}{\mathbf}
\newcommand{\be}{\begin{equation}}
\newcommand{\ba}{\begin{eqnarrray}}
\newcommand{\ee}{\end{equation}}
\newcommand{\ea}{\end{eqnarray}}

\def\part{\partial}

\newcommand{\bmin}{\begin{minipage}{0.495\textwidth}}
\newcommand{\emin}{\end{minipage}}

\newcommand{\Bvec}{\mathbf{B}}

\newcommand{\Avec}{\mathbf{A}}

\begin{document}
\title{Geometric scalings for the electrostatically driven helical plasma state}
\author{Cihan Ak\c{c}ay, John M.~Finn, Richard A.~Nebel, and Daniel C.~Barnes}

\begin{abstract}
A new plasma state has been investigated [C.~Akcay, J.~Finn, R.~Nebel and D.~Barnes, 
Phys.~Plasmas \textbf{24}, 052503 (2017)], with a uniform applied
axial magnetic field in a periodic cylinder of length $L=2\pi R$, driven by helical electrodes. The drive
is single helicity, depending on $m\theta+kz=m\theta-n\zeta$, where $\zeta=z/R$ and $k=-n/R$.
For strong $(m,n)=(1,1)$ drive the state was found to have a strong axial mean 
current density, with a mean-field safety
factor $q_0(r)$ just above the pitch of the electrodes $m/n=1$ in the interior. 
This state has possible applications to DC electrical transformers 
and tailoring of the current profile in tokamaks. We study two geometric issues of interest for these applications: 
(i) scaling of properties with the plasma length or aspect ratio; and (ii) behavior for 
different helicities, specifically $(m,n)=(1,n)$ for $n>1$ and $(m,n)=(2,1)$.


\end{abstract}
\maketitle

The helical state described in 
Ref.~\onlinecite{Akcay2017A} 
(hereafter referred to as AFNB) has cylindrical geometry 
with a uniform applied magnetic field $B_z=B_0$
and single helicity electrostatic drive at the edge $r=r_w$. The results were obtained in a periodic cylinder of length 
$L=2\pi R$, with applied 
electrostatic potential $\phi(r_w)\propto \phi_0 e^{im\theta+ikz}=\phi_0 e^{im\theta-in\zeta}$, 
with $\zeta=z/R$ and $k=-n/R$.  For sufficiently strong drive with $(m,n)=(1,1)-$symmetry,
results featured a steady state with a mean field current density $j_z^{(0,0)}$ and azimuthal magnetic field $B_{\theta}^{(0,0)}$, giving
a quasilinear $q$ profile $q_0(r)=rB_z^{(0,0)}/RB_{\theta}^{(0,0)}\gtrsim 1$ except near the edge, where $q_0(r)$ rises.
These strong drive cases were characterized by Alfv\'{e}nic flows, which drive $B_{\theta}^{(0,0)}$ (and $j_z^{(0,0)}$) by a dynamo or flux conversion mechanism\cite{Bellan-spheromaks,JarboeReview,FinnNebelBathke}, which occurs unless $m=0$ or $k=0$. 
The potential applications of this helically-driven plasma state include DC  electrical transformers and tailoring of the current density in tokamaks. 
As $\phi_0$ increases, $q_0(0)$ decreases toward unity and we define 
$\phi_{min}$ as the smallest value of $\phi_0$ for which $q_0(r=0)\gtrsim 1$. 
For much larger driving voltage, $\phi_0>\phi_{crit}$, a time-asymptotic is no 
longer observed. Effectively, the operational steady-state 
range of the device is 
$\phi_{min}<\phi_0<\phi_{crit}$ with $\Delta \phi_0\equiv\phi_{crit}-\phi_{min}$. Investigations of the transient leading to this helical state, of current density 
streamlines for studying the leakage of current 
from the helical electrodes to the ends, 
and of the sensitivity of these properties to velocity boundary conditions, the resistivity profile 
and the primary electrode shape have been presented in Ref.~\onlinecite{Akcay2017B}.

An important set of issues not addressed in AFNB or in Ref.~\onlinecite{Akcay2017B} relates to how the characteristics 
described there change with the aspect ratio $R/r_w$ (plasma length) and the 
helical symmetry of the drive, \textit{i.e.,} $(m,n)$.
These issues are relevant for the potential application of this helical plasma state as an electrical transformer, 
where its compactness will matter, and where the ability to vary the output electrostatic 
potential relative to the input potential is important. The first issue is closely correlated with cost and both issues are related to the step-up or step-down nature of the transformer.
The results presented here show that the aspect ratio influences the current density in inversely: $I_z \sim r_w/R$, as suggested by 
$q_0(r=0)=rB_z^{(0,0)}/RB_{\theta}^{(0,0)}\gtrsim 1$ for small $r$, i.e.~$j_z^{(0,0)}(r\approx0)\sim B_z^{(0,0)}/Rq_0(0)$.
For driving  by helical electrodes with $m=1$ and $n>1$, we consider two separate cases: in the first, $n$ 
is varied with $R/r_w$ fixed; in the second, $n$ is varied with $R/r_w \propto n$. For the first case. we conclude
that the net current $I_z$ increases with $n$, but the operational range decreases. For the second
case, increasing $n$ effectively concatenates $n$ copies of the $n=1$ state. In principle there is the
possibility of weaker stability properties and therefore a decrease in the operational range. 
However, for moderate values of $n$ this decrease in range is not observed. For driving with $(m,n)=(2,1)$ with a sufficiently large $\phi_0$ there is a region of intermediate radius
$r$ for which $q_0(r)\rightarrow m/n=2$, and as for $m=1$, a range near the wall where $q_0(r)$ is larger. We also find that 
$q_0(r)$ diverges as $r\rightarrow 0$ because of $m=2$ regularity there.

The results in this paper were obtained using the DEBS code\cite{DEBS}, which advances the resistive MHD equations. 
The details are described in AFNB and Ref.~\onlinecite{Akcay2017B}.

We first show investigations of the change in the behavior of the steady state solution as the plasma 
length $L$ or aspect ratio $R/r_w$ is changed, while fixing $(m,n)=(1,1)$. 
Scans at $S=100$ 
have been carried out for $R/r_w= 2, 6, 10, 20, \text{and } 30$ to be compared with the 
nominal value $R/r_w=3$ from AFNB. The velocity boundary conditions are the $E\times B$ conditions of AFNB and of 
Ref.~\onlinecite{Akcay2017B}. 
Note that fixing $n$ while increasing $L$ 
implies that the magnitude of the pitch in the electrodes $\Delta \theta / \Delta \zeta=2\pi /L=-nk=-k$ is weakened.
Results were obtained with four values of the helical potential $\phi_0=0.02$, $0.06$, $0.1$, and $0.2$ 
where the last value corresponds to the nominal value used in AFNB.   
\begin{figure}
\centering
\includegraphics[width=0.5\textwidth]{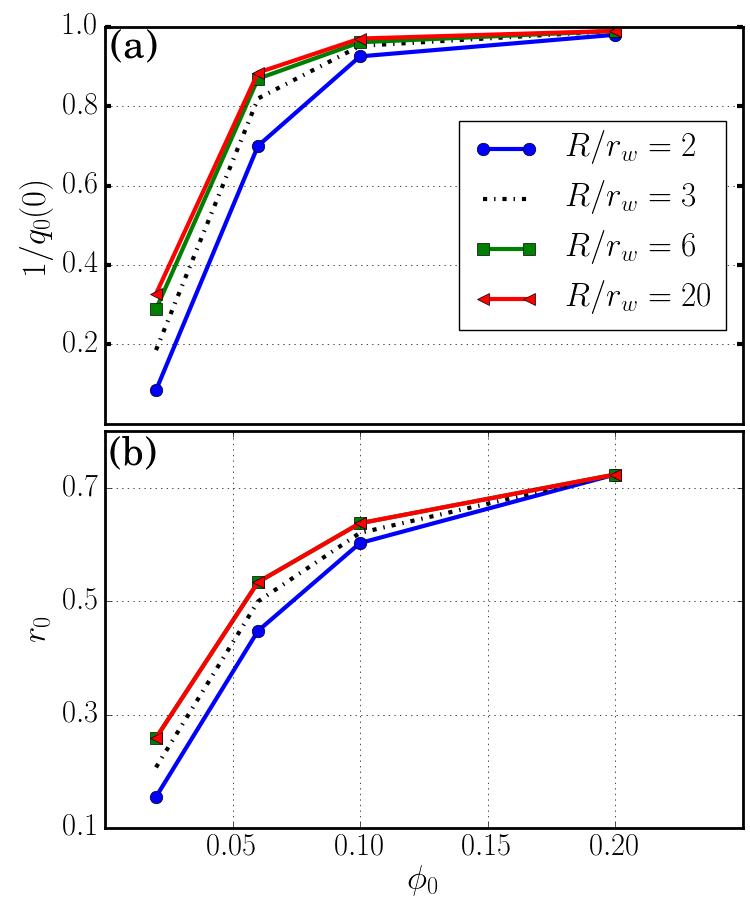}
\caption{\label{fig:q0r0_vs_Phi0_AR} The reciprocal of the mean safety factor on the axis, $1/q_0(0)$, and the 
$O-$point position $r_O$, 
as functions of the applied helical 
potential $\phi_0$ for different aspect ratios $R/r_w$, with $(m,n)=(1,1)$. 
The dependence of these two quantities on $\phi_0$ exhibits little sensitivity to $R/r_w$, above 
$\phi_0=\phi_{min}= 0.1$. }
\end{figure}
\begin{figure}
\centering\includegraphics[width=0.5\textwidth]{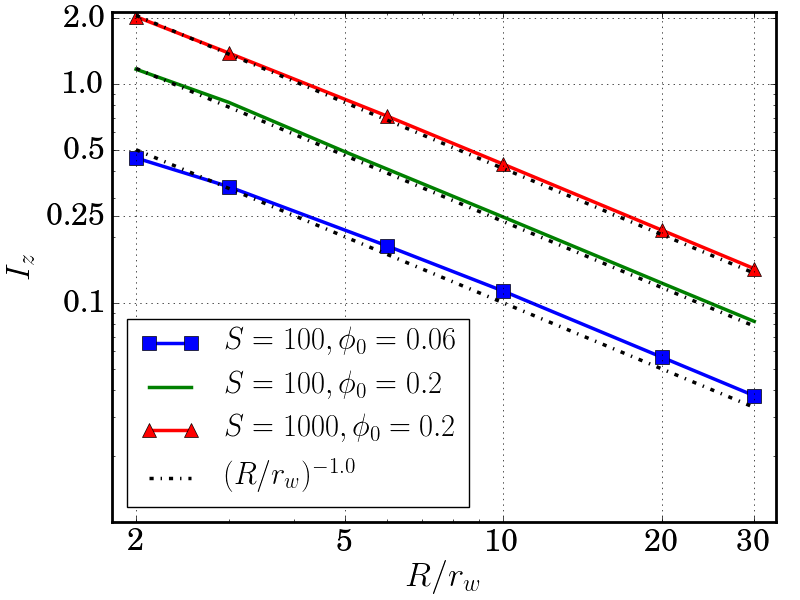}
\caption{\label{fig:Iz_vs_AR} The total current $I_z$ as a function of the aspect ratio $R/r_w$ for $(m,n)=(1,1)$, on a 
log scale, for $\phi_0=0.06$ and $0.2$ at $S=100$ 
(blue squares and solid green) and for $\phi_0=0.2$ at $S=1000$ 
(red triangles). The scans indicate that $I_z$ decreases as $1/R$, as evinced by the black dashed and dotted lines 
that correspond to the 
fits with $1/R$ dependence.}
\end{figure}

The results shown in Fig.~\ref{fig:q0r0_vs_Phi0_AR}a indicate that $\phi_{min}$, above which $q_0(0)\gtrsim 1$, is insensitive to $R/r_w$ for $R/r_w>2$, namely $\phi_{min}\approx 0.1$.
Similarly, the radial position of the elliptic $O-$point shown in Fig.~\ref{fig:q0r0_vs_Phi0_AR}b 
moves in a similar fashion as a function of $\phi_0$, but with $r_O$ increasing more slowly.
The behavior with $r_O$ is well converged 
for $R/r_w\gtrsim 6$ with a noticeable variation for small $R/r_w$. The transition 
to time-dependent behavior, apparently due to an instability as suggested in AFNB, is also observed to occur at nearly the 
same value $\phi_0=\phi_{crit}\approx 0.7$ for all values of $R/r_w$. We conclude that the operational range $\Delta \phi_0$ is insensitive to $R/r_w$ except for small aspect ratios. 

The total plasma current $I_z$ is observed to decrease almost exactly as $1/R$ over this range of $R$, as shown in 
Fig.~\ref{fig:Iz_vs_AR}. 
This scaling holds for higher Lundquist numbers ($S$) as well. 
This is consistent with $q_0(r)\gtrsim 1$, which, with $q_0(0)\gtrsim 1$ and a fixed $B_z^{(0,0)}$, implies 
$j_z^{(0,0)}(0)\propto 1/R$.
The magnetic helicity, discussed at length in AFNB and in 
Ref.~\onlinecite{Akcay2017B}, also
remains insensitive to $R/r_w$ (not shown), in spite of the unwinding of the poloidal field. 
This is because while the integrand $\Avec\cdot\Bvec \propto 1/R$ (because both $A_z$ and $B_{\theta}$ scale as $1/R$), 
the volume is proportional to $R$, thereby canceling the explicit dependence 
on the aspect ratio. Equivalently, as $L=2\pi R$ increases, the poloidal flux increases proportional to $L$ while 
the toroidal flux remains fixed, but their relative linkage decreases at the same rate, because of the unwinding of the 
quasilinear field with $q_0(r)\gtrsim 1$, i.e.~$B_{\theta}^{(0,0)}/B_z^{(0,0)}\sim r/q_0 R$.

The next important issue involves varying the toroidal mode number $n$, with $m=1$, to determine how the time-asymptotic state varies. 
We vary $n$ in two manners: first with $R/r_w$ fixed and then with $R/r_w\sim n$. 
The former amounts to $R=3$ and latter to $R\propto n$ since $r_w=1$ always. 
The scans presented here were run with the $E\times B$ velocity boundary condition (imposed on the 
$(1,1)$ component of $v_r(r_w)$--see AFNB and Ref.~\onlinecite{Akcay2017B}). 
However, additional simulations were also conducted with no-slip and zero-stress velocity boundary conditions of 
Ref.~\onlinecite{Akcay2017B}, yielding the same dependence on $(1,n)$-drive as the $E\times B$ results discussed here. 

The characteristics of the time-asymptotic state with a $(1,n)$ drive and fixed $R/r_w=3$ 
differ from those of the $n=1$ nominal case of AFNB in two major respects: (1) 
narrowing of $\Delta\phi_0$ and (2) the increase of $I_z$ with $n$.

\begin{figure}
\includegraphics[width=0.5\textwidth]{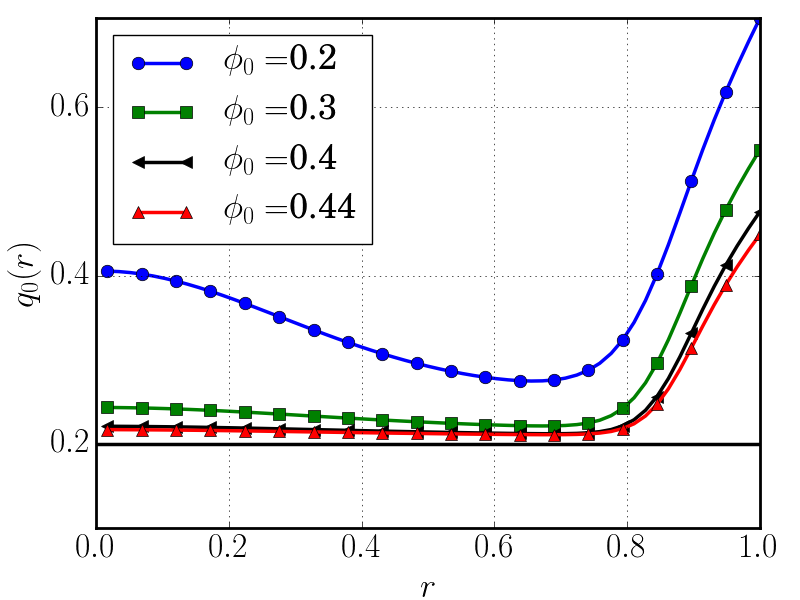}
\caption{\label{fig:q00_vs_r_n5} The mean field profile $q_{0}(r)$ for a case driven with a $(1,5)$ helical electrostatic 
field at various values of the 
applied potential: $\phi_{0}=0.2, 0.3, 0.4$, and $0.44$
for a fixed aspect ratio $R/r_w=3$. The values of $q_0$ approach the solid horizontal line $q_0=1/5$,
but in this case with $q_0(0)$ converging more slowly than $q_0$ located at radii in the middle range.}
\end{figure}

For the toroidal mode $n$ scan with a fixed aspect ratio $R/r_w=3$, the results indicate that $q_0(r=0)\gtrsim 1/n$ for sufficiently strong drive ($\phi_0>0.1$). 
For $n=5$ the sharp transition to $q_0(0)\gtrsim 1/5$ occurs at $\phi_0=\phi_{min}=0.4$ as seen in the $q_0(r)$ profiles in Fig.~\ref{fig:q00_vs_r_n5}. 
The operational range shrinks as $n$ is raised. 
The lower limit $\phi_{min}$ increases with $n$ while the upper limit $\phi_{crit}$ decreases. 
For example, $\phi_{min}=0.1$ for $n=1$ shifts to $\phi_{min}=0.4$ for $n=5$, while $\phi_{crit}=0.70$, $0.54,0.45$ for $n=2,4,5$, respectively.
This decrease is apparently due to the increasing plasma twist (increasing $n$). 
The increased twist enhances the plasma current $I_z$, which approximately scales as $n$, as suggested by $q_0\sim 1/n$. 
Thus, while the cases with $n>1$ and $R$ fixed have a reduced operational range in $\phi_0$, they also yield a greater $I_z$. 
Consequently, magnetic helicity is also greater (not shown) since the twist of the applied field grows approximately as $n-$fold.
We conclude that there is an advantage in using modest $n>1$ with $R/r_w$ fixed, in 
order to cause an increase in the current without too much decrease in operating range $\Delta \phi_0$. 

\begin{figure}
\centering
\includegraphics[width=0.5\textwidth]{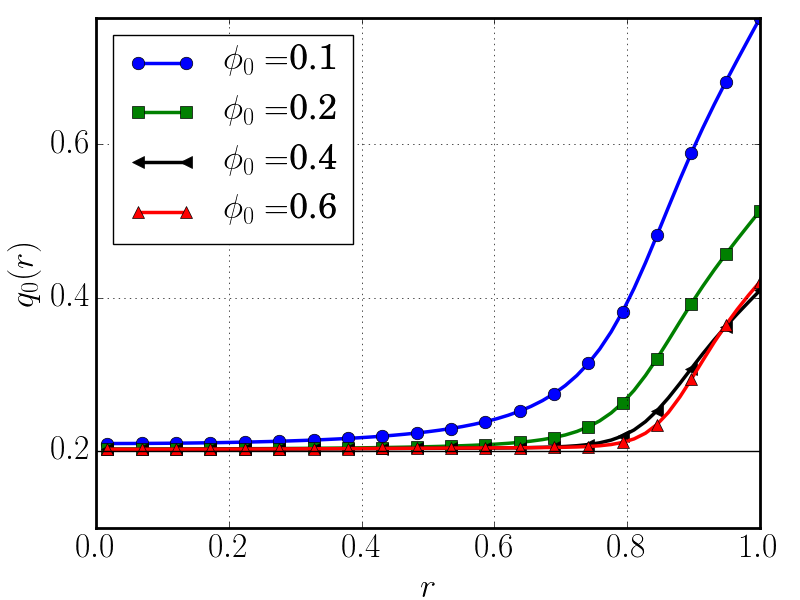}
\caption{\label{fig:q00_vs_r_n5_AR15} The mean field profile $q_{0}(r)$ for a case driven by an 
applied helical electrostatic field with $(m,n)=(1,5)$ 
and four values of the applied potential: $\phi_{0}=0.1, 0.2, 0.4, 0.6$. The aspect ratio is $R/r_w=3n=15$. Note that
$q_0(0)\rightarrow 1/n=0.2$ for large $\phi_0$.}
\end{figure}

The next study consists of increasing $n$ with $R/r_w\propto n$. The profile of the mean field safety factor 
$q_0(r)$ is shown in Fig.~\ref{fig:q00_vs_r_n5_AR15} for four different values 
of $\phi_0$, again for $n=5$, with $R/r_w=3n=15$.
Consistent with AFNB, $q_0(r)$ becomes increasingly flat in the interior, with $q_0(0)\gtrsim 1/n = 1/5$, 
as $\phi_0$ is raised past $\phi_{min}$ 
into the strongly-driven regime. Compare this with the profiles of Fig.~\ref{fig:q00_vs_r_n5}.
The magnetic helicity for $n>1$ (not shown) increases approximately by a factor of $n$ compared to the nominal case because the 
volume of the domain grows as $n$ 
while the overall twist remains the same, \textit{e.g.}~the case with the $(1,5)$ drive has five times as much magnetic helicity as that of AFNB.

The properties of the time-asymptotic state driven with $(1,n)$, while $R\propto n$, are nearly identical to the nominal case of AFNB driven with a $(1,1)$ helical 
mode for $R=3$, with $q_0(0)\rightarrow nq_0(0)$. 
In fact, this scan amounts to attaching end-on 
(in series) $n$-many copies of the nominal $n=1$ state of AFNB. 
It was demonstrated in AFNB that the steady time-asymptotic state is lost beyond a critical applied voltage: 
$\phi_0>\phi_{crit}$. This 
phenomenon is apparently due to instability of the steady state beyond this limit. 
The quantity $\phi_{crit}$ could in principle decrease with $n$, but we have not observed any significant difference, 
\textit{i.e.}~we observe 
$\phi_{crit}\approx 0.7$ for $1\leq n \leq 5$. 
Also, $\phi_{min}$,
the minimum value of $\phi_0$ for which $q_0(0)\gtrsim 1/n$, was observed to remain the same regardless of the value of $n$. 
That is, for the cases with $R\propto n$ the operating range is insensitive to $n$. 
In addition, $I_z$ was also found to be insensitive to $n$, consistent with the behavior $j_z^{(0,0)}= 2/R q_0(0)$ 
with $nq_0(0)\approx 1$ and the fact that the steady state consists of $n$ copies of the $n=1$ state. 

\begin{figure}
\centering
\includegraphics[width=0.5\textwidth]{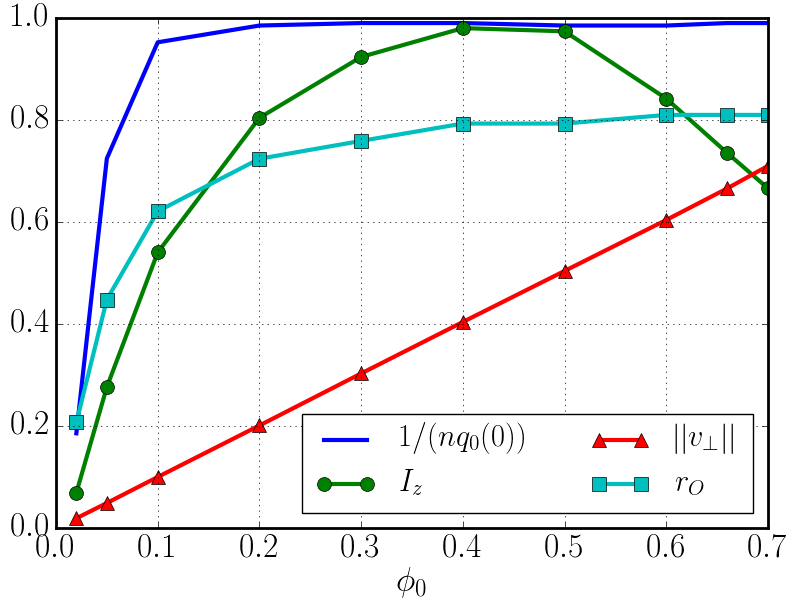}
\caption{\label{fig:n5_metrics} The quantities $1/nq_{0}(0)$ (solid blue), total axial
current $I_{z}$ (green dots), $||v_{\perp}||/v_A$ (red triangles)
and the radial location $r_{O}$ of the $O$-point (cyan squares) as functions of
the amplitude of the applied helical voltage $\phi_{0}$ for $(m,n)=(1,5)$. 
The aspect ratio $R=15$, \textit{i.e.}~scaled 
up from the $n=1$ case as $3n$ to keep the twist per unit length $\propto k$ constant.} 
\end{figure}

Figure \ref{fig:n5_metrics} shows results for $n=5$, identical to results from AFNB (Fig.~3) with $q_0\rightarrow nq_0$. 
This figure clearly indicates $\phi_{min}=0.1$.
The radial position of the $O-$ point of the helical flux $\chi$ shifts monotonically outward 
while the norm of $\mf{v}_{\perp}/v_A$ rises linearly with $\phi_0$ with a slope $\sim 1$. The plasma current $I_z$ has 
a flat peak 
centered around $\phi_0=0.4$. 
These four traces are almost identical to that for $n=1$ drive illustrated in Fig.~3 of AFNB, thus demonstrating that the characteristics are insensitive to the axial mode number if $R/r_w\propto n$, with $q_0\rightarrow nq_0$.

These three sets of $m=1$ scans, the first varying $R/r_w$ with $n=1$ fixed, the second varying $n$ with $R/r_w=3$ fixed,
and the third varying $n$ with $R/r_w=3n$, consist of scans of the two axes and a diagonal in $n-R/r_w$ space.

The studies shown so far, with $m=1$ and varying axial mode number $n>1$, 
while also scaling up the length or toroidal aspect ratio as $n$,
hold up qualitatively in the presence of a back EMF\cite{Akcay2017A}, representative of the resistive load of a secondary circuit. 
Recall $n>1$ behaves as $n$ identical copies of the $n=1$ system. 
Therefore, the back EMF or output voltage for the $n>1$ system simply equals $n$ times the output voltage of for $n=1$: $nE_0 2 \pi R$, where $E_0$ is the $(m,n)=(0,0)$ 
component of the electric field. 
In other words, the output voltage is proportional to the length and a longer device has higher output voltage. 
Furthermore, the plasma length provides a way control the output voltage and therefore the stepup-stepdown nature of the 
transformer. 
We do not pursue back EMF further here because of the bipolar structure of current density $j_z$, which results in strong cancellation of the net current $I_z$ in these studies.
As discussed in AFNB, splitting of the secondary electrodes where $j_z$ changes sign permits the separate extraction of 
positive and negative $j_z$ in two circuits that can be added in parallel or in series. 
A proper treatment of the splitting of the secondary electrodes would require a separate back EMF for the two separate parts of the secondary electrodes, and awaits the model of a finite rather than periodic cylinder. 
We note here that adding the
two circuits in parallel or in series also affects the output voltage and current, and therefore the stepup-stepdown nature.

\begin{figure}
\centering 
\includegraphics[width=0.5\textwidth]{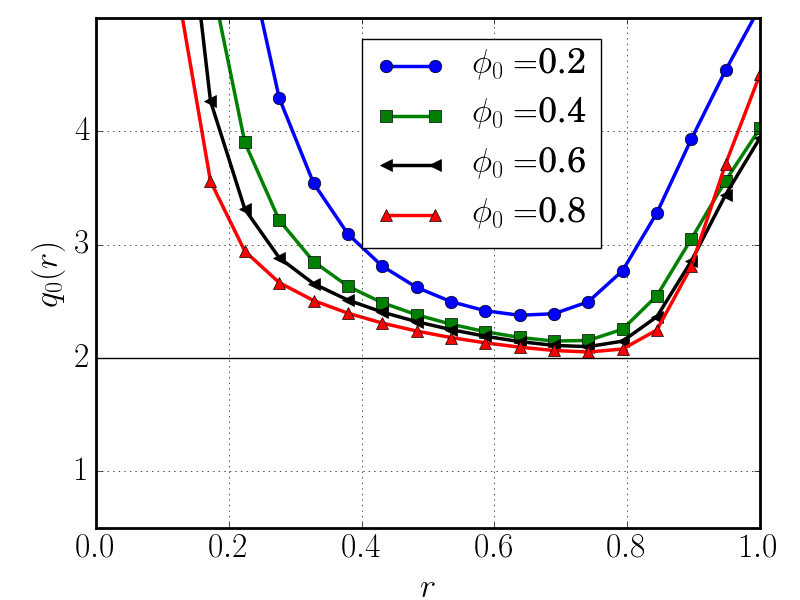}
\caption{\label{fig:q00_vs_r_m2} The mean field profile $q_{0}(r)$ for a case driven with a $(2,1)$ helical electrostatic field at four different values of the applied 
potential, $\phi_{0}=0.2$
(blue circles), $0.4$ (green squares), $0.6$ (solid black), and $0.8$ (red triangles). The solid horizontal line 
marks $q_0=m/m=2$, showing that $q_{min}\rightarrow 2$ as $\phi_0$ increases, but with $q_0\rightarrow \infty$ as 
$r\rightarrow 0$.}
\end{figure}

The steady state found for helicity $(m,n)=(2,1)$ for sufficiently large $\phi_0$ has $q_0(r)$ profiles
shown in Fig.~\ref{fig:q00_vs_r_m2} for various values of $\phi_0$. 
These profiles have minima $q_{min}\gtrsim m/n=2$ in 
the vicinity of $r=0.8$, with $q_{min}$ asymptoting to 2.0 as $\phi_0$ is increased. 
The radial profile of $q_0(r)$ also rises toward the wall similar to cases with $(m,n)=(1,n)$, as in Figs.~\ref{fig:q00_vs_r_n5} and \ref{fig:q00_vs_r_n5_AR15} for $n=5$ and in the nominal $n=1$ state of AFNB. 
The behavior as $r\rightarrow 0$, which leads
to the minimum $q_{min}$, is explained by the regularity conditions at $r=0$ for $m>1$ perturbations. 
The mean field Ohm's law with zero back EMF leads to
\begin{equation}
\langle \hat{\mathbf{z}} \cdot \tilde{\mathbf{v}} \times  \tilde{\mathbf{B}} \rangle_0
=\eta {j}_z^{(0,0)},\label{eq:Dynamo}
\end{equation}
where here $\tilde{\mathbf{v}}$ represents the $(m,1)$ component of $\mathbf{v}$, and similarly for $\tilde{\mathbf{B}}$, 
and $\langle \cdots \rangle_0$ 
represents the average over $\theta$ 
and $z$ with $r$ held fixed, \textit{i.e.}~the $(m,n)=(0,0)$ Fourier component. 
The regularity conditions for $m>1$, which require $\tilde{v}_r \sim \tilde{v}_{\theta} \sim r^{m-1}$ as $r\rightarrow 0$ (and similarly for $\tilde{B}_r$ and $\tilde{B}_{\theta}$) lead to $j_z^{(0,0)}=B_{\theta}^{(0,0)}/r \sim r^{2(m-1)}$ as $r\rightarrow 0$. 
From this, we conclude that the mean field safety factor has $q_0(r) \sim r^{-2(m-1)}$ as $r\rightarrow 0$.
For $m=2$ these results reduce to $j_z(r\sim 0)^{(0,0)}\sim r^{2}$ and $q_0(r\sim 0) \sim r^{-2}$.
Figure \ref{fig:q00_vs_r_m2} also indicates $\phi_{min}\simeq 0.4$ and $\phi_{max}=0.8$, yielding 
a narrower operational range than that of AFNB observed for $(m,n)=(1,1)$. 
Note that both $\phi_{min}$ and $\phi_{max}$ have shifted to larger values for $m=2$.

These results for $m = 2$ suggest possible applications in
which mean current density at the geometric center is not
required. This situation should apply, as the $m = 1$ cases
do, to electrical transformers. However, for the possibility of
driving current in a tokamak, the lack of mean current density at $r = 0$ makes it less likely for $m > 1$ drive to be of use, if the primary reason is to complement the bootstrap current, which is zero there. 
However, it is possible that drive with $m>1$ might be of more general use for tailoring the current density profile.

\begin{figure}
\centering 
\includegraphics[width=0.5\textwidth]{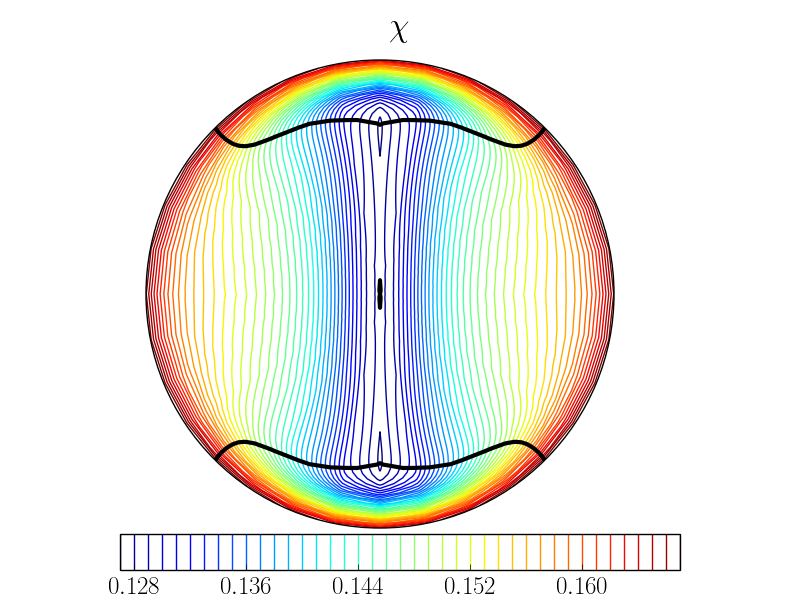}
\caption{\label{fig:ChiJMF} Contours of helical flux $\chi$ for the case $\phi_0=0.2$ of Fig.~\ref{fig:q00_vs_r_m2}. The black curves correspond to $\lambda=0$, which occurs at the origin due to $m=2$ regularity conditions.}
\end{figure}

Figure \ref{fig:ChiJMF} shows the helical flux $\chi$ contours for the $(m,n)=(2,1)$ case of Fig.~\ref{fig:q00_vs_r_m2}
with $\phi_0=0.2$, exhibiting two widely separated $O-$points and an $X-$point on the axis. 
The $\lambda=0$ contours are shown in black, and they indeed connect the X-point
at the center and the two $O-$points. The $\chi$ contours for weaker driving (not shown) have a
single $O-$point and no $X-$point, and also exhibit very elongated surfaces near this limit. 
For stronger driving, the $O-$points are driven more strongly towards the wall and are surrounded by more flux.

We conclude that increasing the length of the device $L=2\pi R$ with $(m,n)=(1,1)$ causes the net current $I_z$ to decrease as
$1/R$. On the other hand, increasing the pitch of the electrostatic drive, \textit{i.e.}~increasing the axial (toroidal) mode number $n$ 
with azimuthal mode number $m=1$ and $R/r_w$ fixed may be useful for the transformer application up to moderate values of $n$, 
because it increases the net current, which in return narrows the operational range. 
Increasing $n$ while simultaneously scaling the aspect ratio by $R/r_w\propto n$ 
essentially creates $n$ copies of the $n=1$ device with no appreciable affect on the operational range, at least up to $n=5$. 
Changing the length or the toroidal mode number $n$
has the potential of changing the ratio of the output voltage to the input voltage, \textit{i.e.}~affecting the step-up / step-down
nature of the transformer. These conclusions apply to the possibility of using a $m=1$ 
electrostatic perturbation to drive current along the magnetic axis in tokamaks. Drive with $m>1$ may be of use in the 
transformer application, but not so much for tokamaks as it cannot drive current down the axis. 

The qualitative results in this paper still hold up in the presence of back EMF, whose detailed study, focusing on the transformer efficiency, is deferred to a future publication.

\section*{Acknowledgements}
We thank Aaron McEvoy, Juan Fernandez, William Gibson, Keith Moser, Liviu Popa-Simil, and Neal Martin for valuable discussions. This research was supported by funding from the ARPA-E agency of the Department of Energy 
under Grant No. DE-AR0000677.


\begin{thebibliography}{1}

\bibitem{Akcay2017A}
C.~Ak\c{c}ay, J.~M. Finn, R.~Nebel, and D.~Barnes.
\newblock Electrostatically driven helical plasma state.
\newblock {\em Physics of Plasmas}, 24(5):052503, 2017.

\bibitem{Bellan-spheromaks}
P.~Bellan.
\newblock Imperial College Press, London, UK, 2000.

\bibitem{JarboeReview}
T.~R. Jarboe.
\newblock {\em Plasma Phys. and Control. Fusion}, 36:945, 1994.

\bibitem{FinnNebelBathke}
J.~M. Finn, R.~A. Nebel, and C.~Bathke.
\newblock {\em Phys. Fluids B}, 4:1262, 1992.

\bibitem{Akcay2017B}
C.~Ak\c{c}ay, J.~M. Finn, R.~Nebel, D.~Barnes, and N.~Martin.
\newblock Properties of the electrostatically driven helical plasma state.
\newblock Technical report, arXiv:1710.03339, 2017.

\bibitem{DEBS}
D.~D. Schnack, D.~C. Barnes~Z. Mikic, D.~S. Harned, and E.~J. Caramana.
\newblock {\em J. Comp. Phys.}, 70:330, 1987.

\end{thebibliography}
\end{document}